# Hyperspectral Nanomotion Microscopy (HNM)

Tongjun Liu [1], Jun-Yu Ou* [1], Kevin F. MacDonald [1], Nikolay I. Zheludev [1, 2]

Corresponding author email: bruce.ou@soton.ac.uk

[1] Optoelectronics Research Centre and Centre for Photonic Metamaterials, University of Southampton, Highfield, Southampton, SO17 1BJ, UK

[2] Centre for Disruptive Photonic Technologies, School of Physical and Mathematical Sciences and The Photonics Institute, Nanyang Technological University, Singapore, 637378, Singapore

**Abstract:** We have developed a technique that extends static scanning electron microscopic imaging to include hyperspectral mapping of fast thermal and externally-driven movements at up to Megahertz frequencies. It is based on spectral analysis of the secondary electron flux generated by a focused electron beam incident on the moving object. We demonstrate detection of nanowire Brownian motion and hyperspectral mapping of stimulated oscillations of flea setae with deep sub-nanometer displacement sensitivity.

Electromagnetic and quantum forces become stronger as the physical dimensions of objects decrease (e.g. the Coulomb force vs. separation between two electrons) and at sub-micron scales, they begin to compete with weakening forces of elasticity (e.g. Hooke's law force vs extension/compression of a spring). Nanoscale movements also become faster as mass decreases with size, reaching Gigahertz frequencies at the nanoscale.[1, 2] These considerations present substantial technological opportunity and explain the growing interest in nanomechanics and the fundamentals of nano- to pico-scale dynamics. Important examples include thermal and stimulated dynamics in micro- and nano-electro-mechanical systems (MEMS & NEMS) such as the accelerometers found in every smartphone, nanomechanical optical metamaterials with variable optical properties controlled by nanoscale movements of artificial "meta molecules", and nanofluidic devices for medical applications.[1-3] Exciting new functionalities emerge through the exploitation of nanomechanical movements of nanoparticles, -powders, -wires, -ribbons, -tubes, -shells, -composites and -membranes (including 2D materials such as graphene); Entropic forces, e.g. Brownian motion, lead to self-organization and cooperative dynamics in nanosystems.[4] Quantum effects, Casimir and Van der Waals interactions play important roles in nanomechanical structures.[5] And over the last decade, protein nanomechanics has seen tremendous progress, evolving into a burgeoning field of biochemical research, while there is considerable interest in adopting bio-inspired solution in advanced materials, nanostructures and bionics.[3,6] However, there are no routinely available imaging technologies that can detect, spatially map and quantify complex, fast nano/picoscale movement of a nano-object.[7] Conventional scanning electron microscopy (SEM) provides nanometre spatial resolution but can operate only at low frame rates, typically up to a few tens per second.[8]

Here we report a nanomotion imaging technique implemented on the platform of a standard scanning electron microscope. When a focused electron beam is scanned over an object, the secondary electron current $I(r)$ is detected and used to generate a static image of the object ($r$ being the coordinate in the object plane). We visualise nanomotion by detecting the time dependence $I(t,r)$ of the secondary electron flux (Fig.1a) at every pixel. In this regime, thermal or small driven displacement of the object $\delta r(t)$ result in secondary electron current changes that are proportional to the gradient of the static image of the object in the displacement direction: $\delta I(t,r) \propto (\nabla I(r) \cdot \delta r(t))$. The technique thereby transduces motion most effectively at the moving edges of an object. A hyperspectral nanomotion microscopy (HNM) image of an object is constructed by mapping the amplitude and phase of spectral components of $\delta I(t,r)$ at every pixel.

HNM can be used to detect thermal (Brownian) motion in the nanostructures, such as the array of nanowires illustrated in Fig. 1b. Here, the incident electron beam is focused to a fixed point at the edge of the central part of each nanowire to maximise the $\delta I(t,r)$ signal. Here, and in all that follows, we employ 5 keV incident electrons at a beam current $I_{in}$ = 86 pA, focused to a spot diameter of ~1 nm. Figure 1c shows thermomechanical displacement spectra for the nanowires at room temperature $T$ with the displacement noise level ~10 picometers (pm) and corresponding displacement sensitivity is ~1 pm/Hz$^{1/2}$. Recorded displacements are found to be in good agreement with thermodynamic estimates $\frac{1}{2\pi f}\sqrt{\frac{k_B T}{m}}$, where $m$, $f$ and $k_B$ are respectively the effective mass, the fundamental oscillation frequency of a nanowire and the Boltzmann constant.

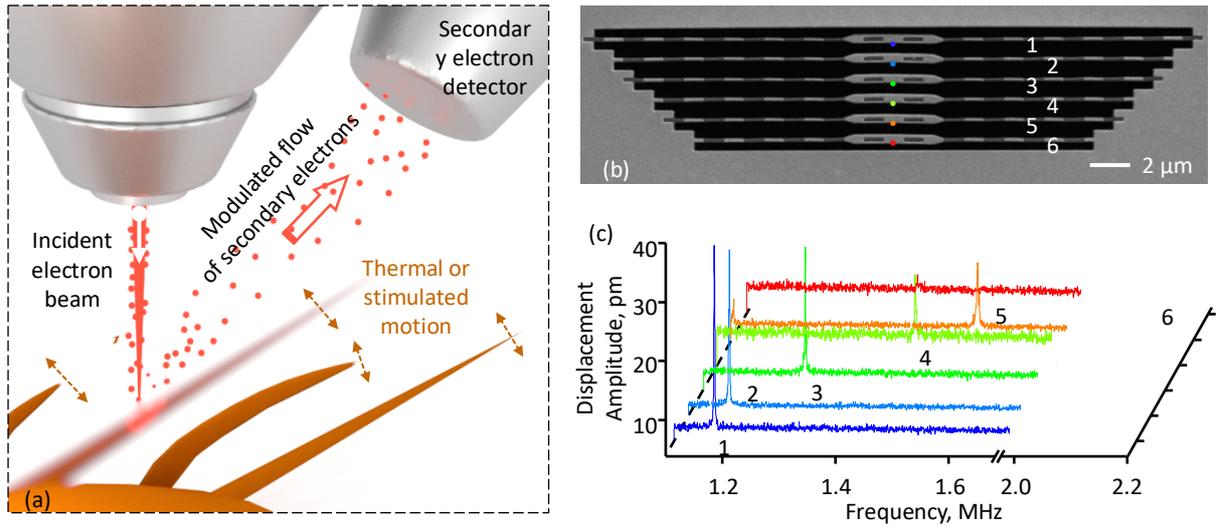

**Fig. 1. Hyperspectral Nanomotion Microscopy.** (a) An incident electron beam generates a flux of secondary electrons that are modulated by fast movements of the target object. Nanomotion images of objects (see Figs. 2 and 3) are constructed by mapping amplitude and phase of spectral components of this secondary electron current at every pixel. (b) An array of six free-standing nanowires of different lengths [from 30 to 20 μm in 2 μm steps] cut into a 100 nm thick gold-coated silicon nitride membrane. (c) Brownian motion spectra of the nanowires present peaks at their length-dependent in-plane resonance frequencies. [Coloured dots in (b) denote positions at which the incident electron beam was focused to record the thermal motion spectra in (c).]

Figure 2 presents HNM images of externally driven motion in the nanowire. For this purpose, the nanostructure was mounted on a piezoelectric actuator oscillating sequentially at the resonant frequencies identified in Fig. 1c while the incident electron beam was raster-scanned over the sample with electron current detected by a lock-in amplifier referenced to the piezo driving frequency. Here we see that the oscillatory movement of individual nanowires can be selectively stimulated, while Figs. 2(f) and (l) show the simultaneous excitation of two wires with overlapping resonance envelopes at 1.171 MHz.

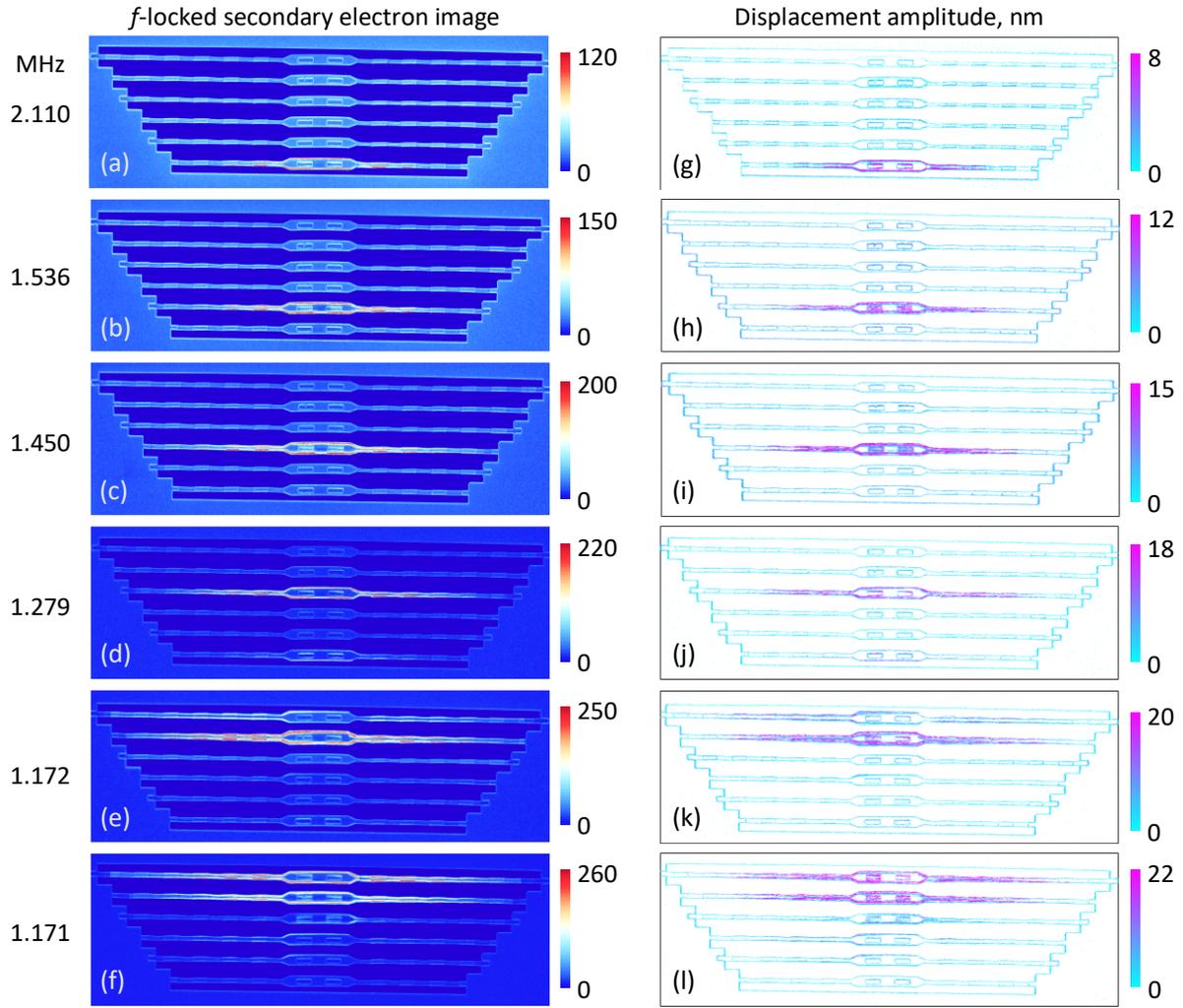

**Fig. 2. Hyperspectral nanomotion microscopy of driven oscillatory motion.** (a) – (f) Maps (arbitrary units) of the amplitude of frequency-locked spectral components [as labelled] of secondary electron current over the nanowire array shown in Fig. 1b. (g) – (l) Corresponding maps induced in-plane displacement amplitude.

Figure 3, referencing Robert Hooke's iconic 1665 microscopic image,[9] shows HNM images of a flea (deceased and mounted on a piezo-actuator, as per the above nanowire sample). The imaged maxillary palpus and coxa region includes setae a few tens of micrometres long and a few hundred nanometres in diameter with natural oscillation frequencies in the MHz range. The HNM technique can uniquely provide data for quantitative evaluation of mechanical properties, in particular elastic moduli, in nanomaterials of such complexity.

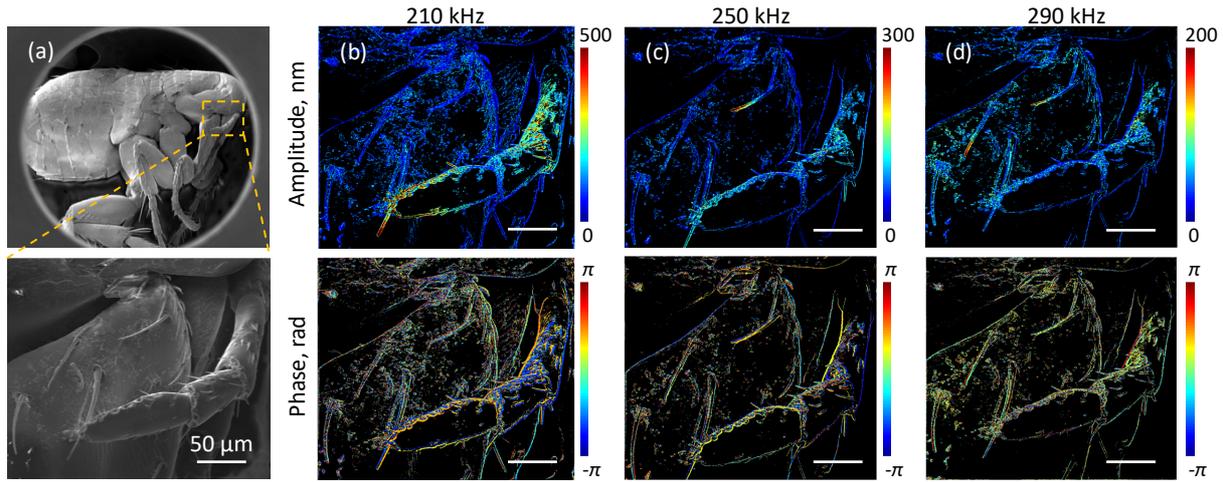

**Fig. 3. Hyperspectral nanomotion microscopy of a flea.** (a) Static SE image of the entire flea and enlarged detail of the HNM imaged area. (b, c, d) frequency-locked amplitude [upper row] and corresponding phase [lower] of secondary electron signal at selected driving frequencies [as labelled].

In summary, we have demonstrated that in electron microscopy the time-domain analysis of secondary electron emission can provide for mapping of thermal and stimulated movement in objects with sub-nanometre displacement resolution. Brownian motion detection provides information on the natural mechanical frequencies of an object, while spatial mapping of coherent oscillations driven by an external force enables the assembly of a hyperspectral nanomechanical portrait of devices and material structures. We argue, that HNM can provide information on the elasticity and mobility of moving parts in biological samples and may be applied to macromolecular structures, the study of protein folding and nanoscale fibre movements leading to cell growth.

## Methods

All experiments were performed on the scanning electron microscope arm of the FEI Helios NanoLab™ 600 DualBeam FIB/SEM system. Acquisition and processing of the signal from the secondary electron detector are performed using Zurich Instruments UHFLI digital lock-in amplifier. For Brownian motion detection, the instrument is configured as a spectrum analyser. In the HNM regime, scanning of the electron beam is controlled by a dual-channel arbitrary function generator (TTi TGF4242) while the lock-in amplifier's internal clock generator is used to drive a piezoelectric sample stage actuator (Thorlabs PA4FEH3W). The lock-in is then referenced to that clock frequency to demodulate the output signal of the secondary electron detector and build amplitude and phase maps of object movements at said frequency. The DC component of the signal is simultaneously employed to build the conventional SEM image of the sample.

**Acknowledgements**

The authors thank Dr Eric Plum for discussions. This work is supported by the EPSRC UK (grant EP/M009122/1) and the MOE Singapore (grant MOE2016-T3-1-006). T. Liu acknowledges and the support of the Chinese Scholarship Council (CSC No. 201806160012).